# NORMAL STATE of the METALLIC HYDROGEN SULFIDE

## N. A. Kudryashov, A. A. Kutukov, E. A. Mazur [a)]

National Research Nuclear University "MEPHI", Kashirskoe sh.31, Moscow 115409, Russia

Generalized theory of the normal properties of the metal in the case of the electron-phonon (EP) systems with not constant density of electronic states is used to examine the normal state of the $SH_3$ and $SH_2$ phase of the hydrogen sulfide at different pressures. The frequency dependence of the real $\operatorname{Re} Z(\omega)$ and imaginary $\operatorname{Im} \Sigma(\omega)$ part of the self-energy part (SP) $\Sigma(\omega)$ of the electron Green's function, the real $\operatorname{Re} Z(\omega)$ and imaginary $\operatorname{Im} \Sigma(\omega)$ part of the complex renormalization of the electron mass, the real $\operatorname{Re} \chi(\omega)$ and imaginary $\operatorname{Im} \chi(\omega)$ part of the complex renormalization of the chemical potential, as well as the renormalized with the strong electron-phonon interaction the electron density of states $N(\varepsilon)$ are calculated. The calculations are performed for the stable orthorhombic IM-3M hydrogen sulfide $SH_3$ structure at three pressures P = 170 GPa, P = 180 GPa, P = 225 GPa and $SH_2$ structure with the symmetry *I4/MMM (D4H-17)* for the three values of pressure P = 150 GPa, P = 180 GPa, P = 225 GPa at the temperature of T = 200K.
Key words: hydrogen sulfide, electron-phonon system, the pressure, the electronic properties.

*e-mail: eugen_mazur@mail.ru*

## INTRODUCTION

In [1,2] it was discovered the superconducting transition of the metal sulfide phase under a pressure of about 170 GPa at T = 203K. The high value of $T_c$ in the hydrogen sulfide under pressure [1,2] is a property of only the electron-phonon system. The presence of both the isotope effect and the Meissner effect in the hydrogen sulphide phase formed at a pressure of about 170GPa has been experimentally demonstrated in [2]. In [3] the main candidate for the superconducting phase namely the compound with the stoichiometry of $SH_3$ and IM-3M symmetry is examined. In [4] along with *$SH_n$* (n≤3) compounds the $SH_2$ phase was found having symmetry I4 / MMM, for which the electron and phonon spectra are calculated, as well as the density of electron and phonon states in the pressure range of 100-225 GPa. In [5-9] it is shown that the main candidates for the role of the superconducting phase are the compounds with the stoichiometry of $SH_3$ and $SH_2$.

Quantum theory of the electronic properties of the normal state of the crystal, as well as the Éliashberg theory of the superconducting state, both are based on the work of Migdal [10]. When considering the properties of the hydrogen sulfide EP system it is critical to take into account the fact of abrupt changes in the density of electron states at the energy scale comparable with the characteristic phonon energy, as well as the finiteness of the electronic band gap. Apparently, the hydrogen sulfide under a pressure is an ideal material for the manifestation of this effect of the reconstruction of the electron bands. This is due to the fact that in a metal hydrogen sulfide occur both the coincidence and the imposition of the role of three key factors mamely: 1. finite width of the conduction band at the same time with the sharp changes in the electronic density of states in this band; 2. very high phonon energy $\hbar\omega_{ph} \sim 0.25 eV$ associated with the collective oscillations of quasi-two dimensional metallic hydrogen, stabilized with the sulfur sublattice; and 3. greater than one constant $\lambda$ of the electron-phonon coupling. The exceptionally strong electron-phonon coupling in the metallic hydrogen sulfide occurs due to the absence of the electron shells in the protons. It is because of these three factors in a metal hydrogen sulfide one should expect a fundamental restructuring of the electron spectrum with the strong electron-phonon interaction.

2. ELECTRONIC PROPERTIES OF $SH_3$ AND $SH_2$ HYDROGEN SULFIDE PHASES, NOT RENORMALIZED WITH THE ELECTRON-PHONON INTERACTION

The hydrogen sulfide $SH_3$ structure (Fig.1, [4]) is considered [3-9] as one of the two main contenders for the phase responsible for the superconductivity effect in the experiments [1,2], along with coexisting with $SH_3$ phase a metastable $SH_2$ phase.

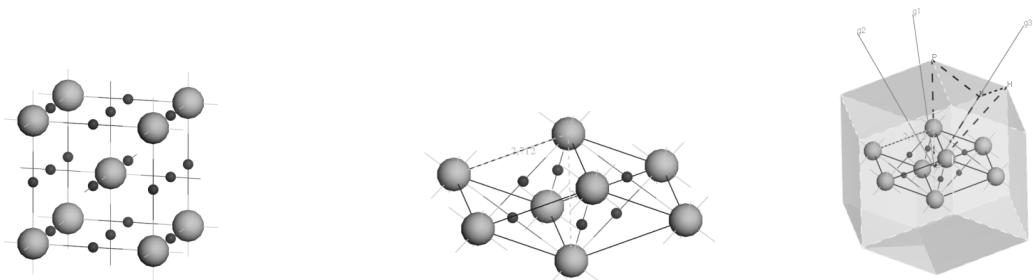

| *a* | *b* | *c* |
|---|---|---|

Fig.1. The SH$_3$ structure in study. The sulfur atoms are presented by a larger size balls; the results of the calculations at a pressure P = 150GPa are presented; a - the initial cubic cell; b - the primitive cell with IM-3M (OH9) symmetry; c - Brillouin zone corresponding to the primitive cell.

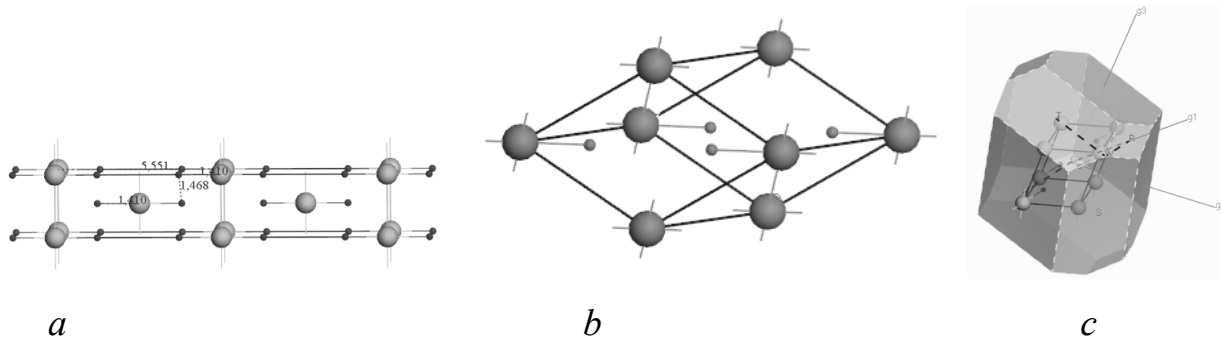

| *a* | *b* | *c* |
|---|---|---|

Fig.2. The SH$_2$ structure under investigation. Sulfur atoms are represented by the larger size balls; the results of the calculations at the pressure P = 170GPa; a - the initial cubic cell; b - a primitive cell with the I4 / MMM (D4H-17) symmetry; c – Brillouin zone corresponding to the primitive cell. An image from [4] is given.

In [4] the calculations were carried out for the electron and phonon properties of t SH$_3$ phase at a pressure of P = 225 GPa. The square unit cell of this structure has the IM-3M symmetry. It was shown that for the density of electron states for the SH$_3$ phase at a pressure P = 225 GPa the value of the Fermi level is always located at the height of the electronic DOS. The energy band structure of electrons in the SH$_3$ phase, as well as the density of electron states in the SH$_3$ phase presented in [5–9] are the band structures that take into account the ideal crystal lattice symmetry IM-3M of the SH$_3$ phase as well as the the electron-electron interaction with allowance for exchange and correlation effects. However, the calculated band data for the SH$_3$ phase with the IM-3M symmetry as well as for the SH$_2$ phase with the I4 / MMM (D4H-17) symmetry does not take into account the effects of the renormalization of the electron spectrum of the band with the strong electron-phonon interaction.

# 3. THE RENORMALIZATION OF THE ELECTRON SPECTRUM WITH THE ELECTRON-PHONON INTERACTION

We will carry out the description of the effects of the renormalization of the electron spectrum in a metal hydrogen sulfide with phonons with the quantum field methods using self-consistently the effects of the restructuring of the complex self-energy part $\Sigma(\omega)$ of the electron Green's function in the hydrogen sulphide together with the restructuring of the electron density of states. For the imaginary component $\operatorname{Im}\Sigma(\omega) = -\operatorname{Im}Z(\omega)\omega + \operatorname{Im}\chi(\omega)$ of the electron Green's function self–energy part in the [11-13] the following expression is obtained:

$$\operatorname{Im}\Sigma(\omega) = -\pi \int_0^{+\infty} dz \alpha^2(z) F(z) \times \\ \times \{[N(\omega-z)+N(\omega+z)]n_B(z) + N(\omega-z)f(z-\omega) + N(\omega+z)f(z+\omega)\}. \quad (1)$$

The expression for the real part $\operatorname{Re}\Sigma(\omega) = [1-\operatorname{Re}Z(\omega)]\omega + \operatorname{Re}\chi(\omega)$ of the GF self–energy part has the following form [11-13]:

$$\operatorname{Re}\Sigma(\omega) = -P \int_0^{+\infty} dz \alpha^2(z) F(z) \int_0^{+\infty} dz' \{f(-z') \times \\ \times \left(-\frac{N(-z')}{z'+z+\omega} + \frac{N(z')}{z'+z-\omega}\right) + f(z')\left(-\frac{N(-z')}{z'-z+\omega} + \frac{N(z')}{z'-z-\omega}\right)\}, \quad (2)$$

where $n_B(z)$ is the Bose distribution function, $f(z')$ is the Fermi distribution function, $Z(\omega)$ is the complex renormalization of the electron mass, $\chi(\omega)$ is the quantity, commonly called the renormalization of the chemical potential. In (1) and (2) the renormalized with the EP interaction electronic density of states $N(z')$ is expressed through the "bare" electron density of states $N_0(\xi)$

$$N(z') = -\frac{1}{\pi} \int_{-\mu}^{\infty} d\xi' N_0(\xi') \operatorname{Im} g_R(\xi', z'). \quad (3)$$

This density of states is not symmetrical (even) $z'$ function. In (3) the $\operatorname{Im} g_R$ expression is given by the following formula

$$\operatorname{Im} g_R(\xi,\varepsilon) = \frac{\operatorname{Im}\Sigma(\varepsilon)}{[\varepsilon - \xi - \operatorname{Re}\Sigma(\varepsilon)]^2 + [\operatorname{Im}\Sigma(\varepsilon)]^2}, \quad (4)$$

so that (1) - (4) is a non-linear system. Formula (1-4) take into account the frequency dependence of the electron density of states $N_0(\xi)$, as well as the effect of the finiteness of the electronic band gap. In (3) $g_R$ is a retarded e-GF, $\alpha^2(z)F(z)$ is the spectral function of the EP interaction, $N_0(\xi)$ is a "bare" (not renormalized with the EP interaction) variable electronic density of states defined by the following expression $\int_{S(\xi)} \frac{d^2 p'}{v_{\xi p'}} d\xi = \int_{S(\xi)} N_0(\xi, \omega') d\xi$ with the energy $\xi$ of the "bare" electrons. A record (1) - (4) assumes an average over all directions of the pulse corresponding to the energy surface $\xi$. The replacing of a detailed description of the electron dispersion law on the average characteristics such as the density of electronic states seems to be a good approximation, which allows you to enjoy all of the most important features of the behavior of the electron-phonon system. The recorded system of equations "saves" the total number of electrons, because, along with the reconstruction of bands simultaneously describes the renormalization of the real part of the chemical potential, corresponding to the shift of the Fermi level by the EP interaction. When writing the system of equations (1) - (4) we started from the justice of the "Migdal Theorem" [10], according to which the vertex corrections because of the additional phonon lines can be neglected, since the value of these amendments is small up to the terms defined by the varying degrees of the small parameter $\lambda \hbar \omega_D / E_{cond} \ll 1$, where $E_{cond}$ is the width of the conduction band. In the formulation of the mathematical model described by equations (1) - (4), we also neglected multi-phonon vertices. The accounting for such multi-phonon vertices leads to the contributions corresponding to the interaction of an electron simultaneously with two or more phonons. We have taken into account that the constant of the simultaneous electron interaction with a pair or more of phonons is extremely small [14−16]. At the same time the consecutive interaction of electrons with phonons, corresponding to the contributions from the accounting one-phonon vertex in any order on the powers of the electron-phonon interaction in this study is completely taken into account. Indeed, solving of equations (1) - (4) with the analytical method of successive approximations and expanding the solution in powers of the integral of the spectral function of the EP interaction $\alpha^2(z)F(z)$ (in other words, in the powers of the constant $\lambda$ of the EP interaction), we see that $\text{Re}\Sigma(\omega)$ as well as $\text{Im}\Sigma(\omega)$ contain an infinite number of deposits corresponding to the serial interaction of electrons with phonons randomly. When writing the system (1) - (4) the renormalization of the phonon Green's function with the EP interaction has not been evaluated. As is shown in [17], the holding such a renormalization leads to the double counting of the EP interaction contributions to the phonon Green's function and, consequently, to the appearance of the actually absent negative phonon frequency, which corresponds to a false lattice instability at the significant EP interaction.

4. THE METHODS OF CALCULATION AND RESULTS

Substitute "bare", not renormalized with the electron-phonon interaction, electron density of states for the hydrogen sulfide obtained in the works [3,4], with the dimensionless argument, expressed in fractions of the maximum phonon frequency as well as the pressure dependent function of the electron-phonon interaction $\alpha^2(z)F(z)$ (Fig. 3, 4) with the dimensionless argument, also expressed as a fraction of the maximum phonon frequency to the nonlinear system of integral equations (1) - (4), describing the normal metal hydrogen sulfide. The maximum phonon frequency amounts for the hydrogen sulfide $SH_3$ phase to 0.214 eV, and for the hydrogen sulfide $SH_2$ phase amounts to 0.275 eV. The "bare" electronic density of states does not contain in their behavior "failures" to zero values at the energies below the Fermi level, despite having completely filled bands lying below the energy of the conduction band. This follows from the fact that although these bands overlap each other in the energy, in the Brillouin zone these bands are located in different areas of the momentum space. The main contribution to the density of electronic states in the conduction band makes energy argument values lying within the range of -4 to +4 dimensionless units of energy, that is, within the energy interval of one electron volt both up and down from the Fermi level. The fact of the repeated intersection of the conduction band and the Fermi surface, clearly visible in [4], results in the presence of multiple peaks and minima at the Fermi surface, and hence the Van Hove singularities in the density of electronic states behavior for the $SH_3$ phase. Small contribution to the density of states near the Fermi level of the other areas, which are located partly at the Fermi surface will be neglected. At the energies below the Fermi level the overlap of the conduction band with the underlying zone is negligible. The system of equations (1) - (4) describing the normal state electronic structure of the $SH_3$ and $SH_2$ phases of the hydrogen sulfide interacting with a strong EP interaction was solved at different pressures at T = 200 K by the method of successive approximations to achieve the self-consistency effect. Spectral functions of the electron-phonon interaction $\alpha^2(z)F(z)$ for the test phases $SH_3$ и $SH_2$ of the hydrogen sulfide presented in Fig. 3, 4, are similar to that of the phonon density of states in the sulfide phase, which coincides with the findings [5-9]. In this situation the EP interaction constant equals to $\lambda = 2\int_0^\infty d\omega \frac{\alpha^2(\omega)F(\omega)}{\omega} \approx 2.2$ for the $SH_3$ phase and equals to $\lambda \approx 1.0$ for the $SH_2$ phase. Such values of the EP interaction constants indicate the strong EP interaction in the hydrogen sulfide phases in the study.

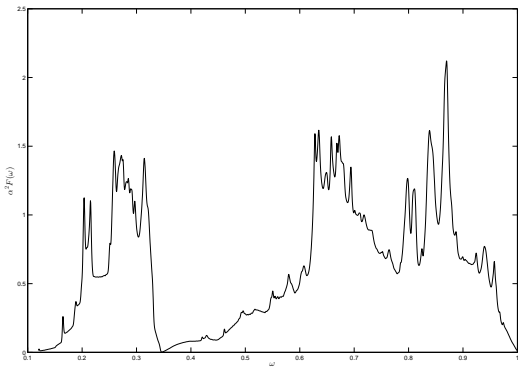

Fig. 3. The spectral function of the electron-phonon interaction $\alpha^2(\omega)F(\omega)$ in the $SH_3$ hydrogen sulfide at a pressure of 225 GPa. The frequency $\omega$ is expressed in dimensionless units (as a fraction of the maximum frequency of the phonon spectrum, equal to 0.214 eV).

The main features of the spectral function of the electron-phonon interaction $\alpha^2(\omega)F(\omega)$ in the $SH_3$ hydrogen sulfide at a pressure of 225 GPa presented in Fig.

3, as well as the corresponding constant of the electron-phonon interaction $\lambda \approx 2.2$ coincide with the similar characteristics from $[5-9]$.

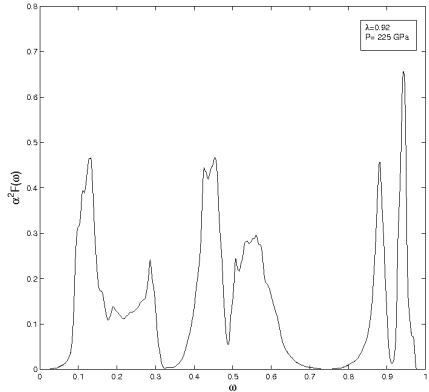

Fig. 4. The spectral function of the electron-phonon interaction $\alpha^2(\omega)F(\omega)$ for the $SH_2$ hydrogen sulfide at a pressure of 225 GPa ($\lambda = 0.92$). The frequency $\omega$ is expressed in dimensionless units (as a fraction of the maximum frequency of the phonon spectrum, equal to 0.275 eV).

To solve the nonlinear system of integral equations (1) - (4) the program of numerical simulation of the electronic density of states was written. The convergence of computing was achieved after about the tenth iteration for a given choice of computing step about one hundredth. The convergence of the iterative process depends on the computation step. It was found that a change in the calculation step two or three times does not significantly affect the number of iterations to produce the final picture of the electronic density of states. Characteristic factor is the dependence in the calculations carried out of the convergence rate of the iterative process on increasing the computational domain for the density of electronic states.

5. NORMAL PROPERTIES OF THE $SH_3$ HYDROGEN SULFIDE PHASE

Fig. 5 shows the results of solving the system of equations (1) - (4) for the real part $\mathrm{Re}\Sigma(\omega)$ and an imaginary part $\mathrm{Im}\Sigma(\omega)$ of the self-energy part of the electron Green's function of the hydrogen sulfide $SH_3$ phase in the range of dimensionless energy variable from -10 to +10, which corresponds to the most interesting with respect to the superconductivity energy region about -2.14eV to +2.14 eV. Fig. 6 shows a part of the reconstructed conduction band (right panel) at different spectral functions of the electron-phonon interaction, corresponding to different pressures, and as a result, various $\lambda$ parameter values, in comparison with the initial "bare" density of electronic states.

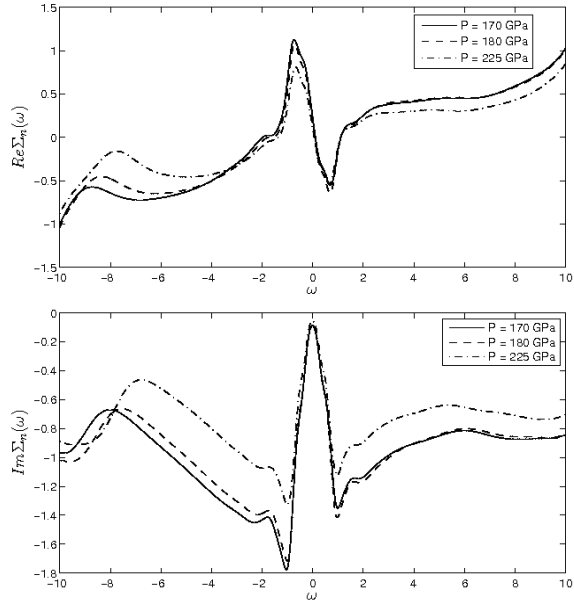

Fig. 5. a. The real part $\text{Re}\Sigma(\omega)$ of the self-energy part of the electron Green's function in the $SH_3$ phase of the hydrogen sulphide, b. The imaginary part $\text{Im}\Sigma(\omega)$ of the self-energy part of the electron Green's function. Both graphs are shown in the range of 2.14 electron volts at both sides of the Fermi level. Frequency $\omega$ is expressed in dimensionless units (as a fraction of the maximum frequency component 0.214 eV of the phonon spectrum for this hydrogen sulfide phase). The results are obtained for the following pressure values: P = 170 GPa ($\lambda$ = 2.599), P = 180 GPa ($\lambda$ = 2.589), P = 225 GPa ($\lambda$ = 2.273) and the temperature T = 200K.

From Fig. 6 it is clear that the part of the conductivity band in the hydrogen sulphide, away from the Fermi level on the order of an electron volt, is experiencing reconstruction under the influence of multiple strong interaction with the phonons. The most interesting for the possible effect of superconductivity rebuilt part of the conduction band is located within the energy intervals of a few units of the characteristic phonon energy from the Fermi energy.

Fig. 6. Dimensionless total density of electronic states of the hydrogen sulfide $SH_3$ phase. Left: "bare" electron density of states in the conduction band in the SH3 hydrogen sulfide phase. To the right: reconstructed with the EP interaction density of electronic states with different values of the electron-phonon interaction constant, corresponding to the three values of pressure. All results were obtained for the three pressure values, namely: P = 170 GPa ($\lambda$ = 2.599), P = 180 GPa ($\lambda$ = 2.589), P = 225 GPa ($\lambda$ = 2.273) and the temperature T = 200K. The frequency $\omega$ is expressed in the dimensionless units (as a fraction of the maximum frequency component 0.214 eV of the phonon spectrum for this hydrogen sulfide phase).

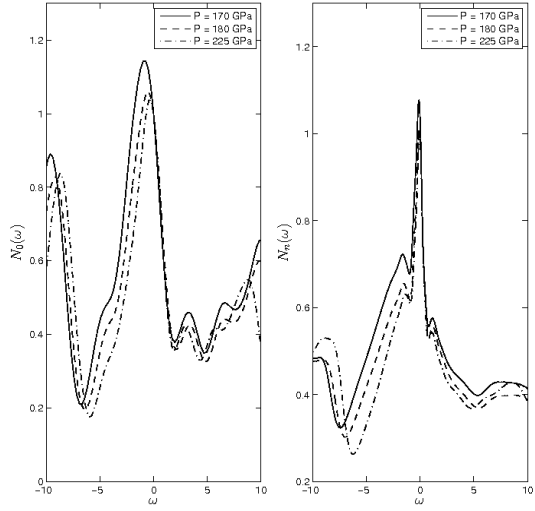

These calculations fully reproduce the polaron effects in multiple and strong interaction of electrons with phonons. Fig. 7 shows the results of the calculation for the $SH_3$ phase for the $\mathrm{Re}\,Z(\omega)$, $\mathrm{Im}\,Z(\omega)$ quantities of the hydrogen sulfide, which describe the real and imaginary parts of the renormalized mass of the electron, as well as estimates for the renormalized with the strong interaction with phonons the real $\mathrm{Re}\,\chi(\omega)$ and imaginary $\mathrm{Im}\,\chi(\omega)$ parts of the $\chi(\omega)$ value for the energy "distances" away from the Fermi level not exceeding 10 characteristic phonon energies, received per unit. The behavior of the electron density of states in this energy range responds to the presence of the major contributions to the density of states in the reconstructed conduction band and is crucial for the existence of the superconducting properties of the material.

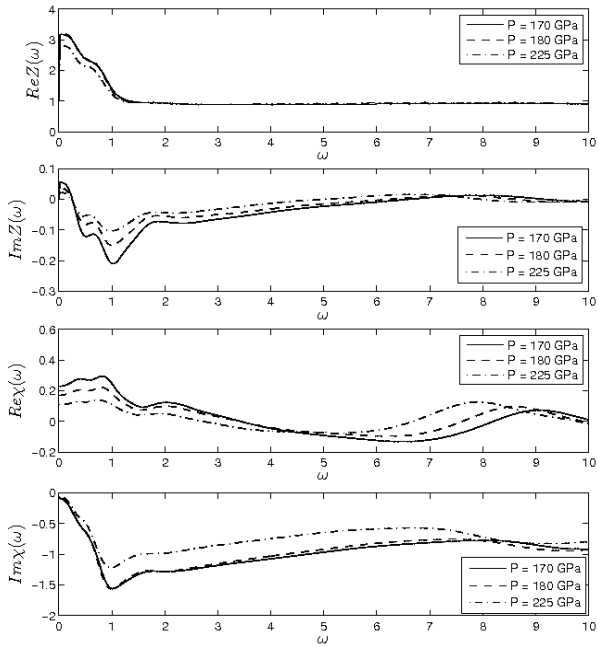

Fig. 7. Reconstructed band parameters of the metal $SH_3$ sulfide phase conduction band. a. The real part $\mathrm{Re}\,Z(\omega)$ of the electron mass renormalization $Z(\omega)$ of the Green function in the $SH_3$ phase of the hydrogen sulphide; b. The imaginary part $\mathrm{Im}\,Z(\omega)$ of the renormalization of the electron mass inside the self-energy part of the electron Green functions; c. Renormalized with the electron-phonon interaction the real part $\mathrm{Re}\,\chi(\omega)$ of the renormalization of the chemical potential of hydrogen sulfide in the $SH_3$ phase. d. Renormalized with the electron-phonon interaction the imaginary part $\mathrm{Im}\,\chi(\omega)$ of the renormalization of the chemical potential of hydrogen sulfide in the $SH_3$ phase. Frequency $\omega$ is expressed in dimensionless units (as a fraction of the maximum frequency component 0.214 eV of the phonon spectrum for this hydrogen sulfide phase). All results are obtained for the three

pressure values, namely P = 170 GPa (λ = 2.599), P = 180 GPa (λ = 2.589), P = 225 GPa (λ = 2.273) and the temperature T = 200K.

On the Fig. 7, basing on the fact of the $\text{Im}Z(\omega)$ sign-alternating behavior at low frequencies $\omega < 2$, one can observe the effect of the reconstruction of the part of the conduction band, adjacent to the Fermi level, to a series of energy-tight "pockets" without overlapping on the energy variable. A set of "pockets", which are located at the energy distances of the order unity from the Fermi level is the most important for the superconductivity part of the conduction band, reconstructed with the strong electron-phonon interaction. Figure 8 shows the graphs of the real part $\text{Re}\Sigma(\omega)$ of the self-energy part of the electron Green's function in the $SH_3$ phase of the hydrogen sulfide, the imaginary part $\text{Im}\Sigma(\omega)$ of the self-energy part of the electron Green's function, and the renormalized with phonons the density of electron states $N(\omega)$ of the conduction band in a broader range of -20 to 20 dimensionless frequency $\omega$.

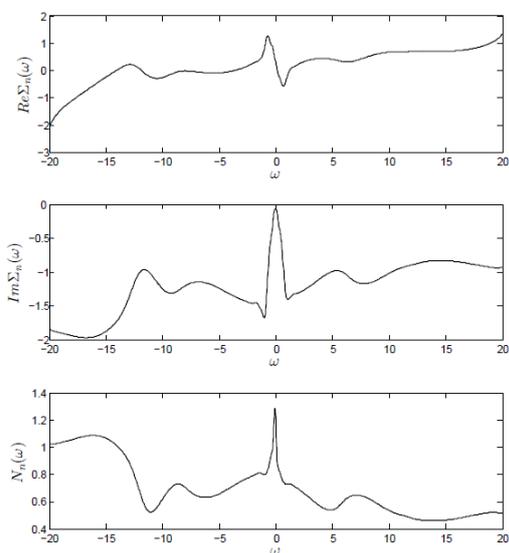

Fig. 8. a. The real part $\text{Re}\Sigma(\omega)$ of the self-energy part of the electron Green's function in the $SH_3$ phase of the hydrogen sulphide. b. The imaginary part $\text{Im}\Sigma(\omega)$ of the self-energy part of the electron Green function. c. The renormalized with the electron-phonon interaction density of electron states $N(\omega)$ in the hydrogen sulphide phase in the range of 4.28 electron-volts on both sides of the Fermi level. Frequency $\omega$ is expressed in the dimensionless units ( the highest phonon frequency spectrum component for this hydrogen sulfide phase amounts to 0.214 eV). All the results are obtained for the pressure P = 225 GPa (λ = 2.273) and the temperature T = 200K.

6. NORMAL PROPERTIES OF the $SH_2$ HYDROGEN SULFIDE PHASE.

Fig. 9 shows the results of solving the system of equations (1) - (4) for the metal $SH_2$ sulfide phase for the real component $\text{Re}\Sigma(\omega)$ and for the imaginary component $Im\Sigma(\omega)$ of the self-energy of the electron Green's function in a range of the dimensionless energy variable from -10 to +10, which corresponds to the most interesting in relation to the superconductivity energy region about -2.75eV to +2.75 eV.

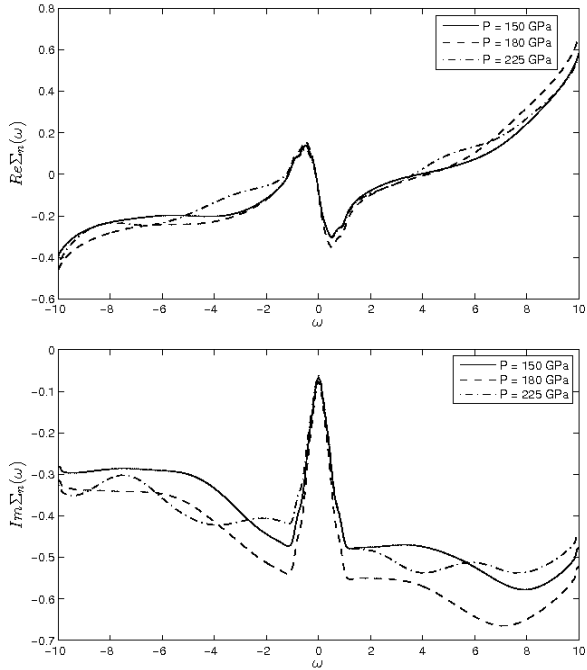

Fig. 9. a. The real part $Re\Sigma(\omega)$ of the self-energy part of the electron Green functions in the hydrogen sulphide $SH_2$ phase; b. The imaginary part $Im\Sigma(\omega)$ of the self-energy part of the electron Green's function. Both graphs are shown in the range of 2.14 electron volts at both sides of the Fermi level. Frequency $\omega$ is expressed in the dimensionless units (as a fraction of the maximum frequency component of the phonon spectrum for this hydrogen sulfide phase which is equal 0.275 eV). The results obtained for the three pressure values P = 150 GPa ($\lambda$ = 1.001), P = 180 GPa ($\lambda$ = 1.11), P = 225 GPa ($\lambda$ = 0.92) and the temperature T = 200K.

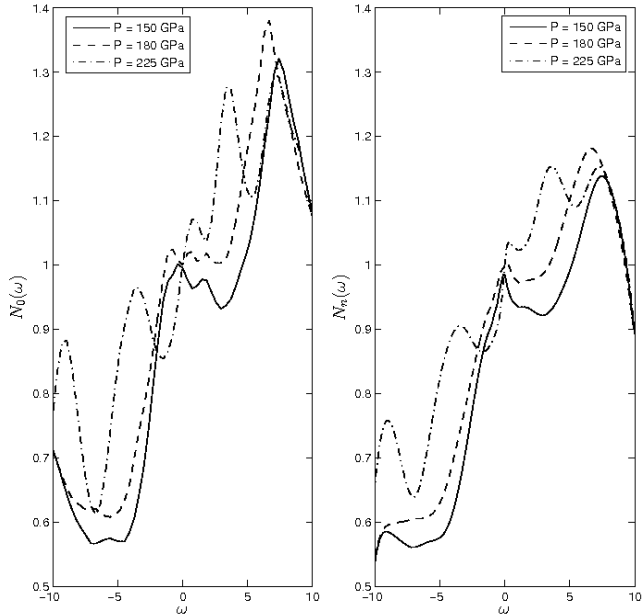

Fig. 10 presents the most important for the phenomenon of superconductivity part of the reconstructed conduction band (right side of Fig. 10) at various spectral functions of the electron-phonon interaction, corresponding to different values of the parameter $\lambda$, in comparison with the initial "bare" density of electronic states (the left side of Fig. 10). From Fig. 10 it is clear that the part of the conduction band in the hydrogen sulphide, away from the Fermi level on the order of an electron volt, is experiencing the reconstruction under the influence of multiple strong interaction with phonons.

Fig. 10. Dimensionless total density of electronic states in the hydrogen sulfide $SH_2$ phase. Frequency $\omega$ is expressed in dimensionless units (as a fraction of the maximum frequency of the phonon spectrum, equal to 275 meV). Left- "bare" electron density of states in the conduction band in the phase of the $SH_2$ of hydrogen sulfide; right: reconstructed with the EP interaction density of electronic states with different values of the electron-phonon interaction, corresponding to the three values of pressure. All results were obtained for the three pressure values P = 150 GPa ($\lambda$ = 1.001), P = 180 GPa ($\lambda$ = 1.11), P = 225 GPa ($\lambda$ = 0.92) and the temperature T = 200K.

Fig. 11 shows the detailed results of the calculation for the $SH_2$ hydrogen sulfide phase for the $\text{Re}Z(\omega)$, $\text{Im}Z(\omega)$ values, which describe the real and imaginary parts of the renormalized mass of the electron, as well as the estimates for the renormalized with the strong interaction with phonons the real $\text{Re}\chi(\omega)$ and imaginary $\text{Im}\chi(\omega)$ parts of the $\chi(\omega)$ value near the Fermi level in the energy "distances" not exceeding 10 characteristic phonon energies, taken per unit.

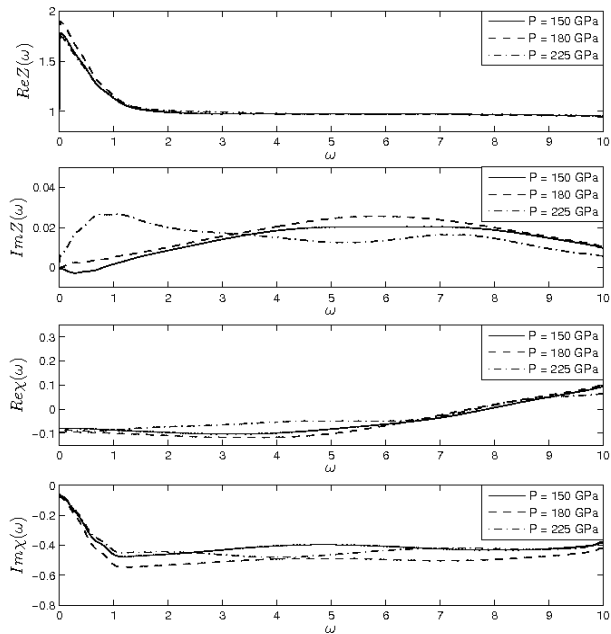

Fig. 11. Reconstructed band options of the metal SH2 sulfide conduction band. a. The real part $\text{Re}Z(\omega)$ of the electron mass renormalization $Z(\omega)$ of the electron Green function; b. The imaginary part $\text{Im}Z(\omega)$ of the renormalization of the electron mass of the self-energy part of the electron Green's function; c. Renormalized with the electron-phonon interaction real part $\text{Re}\chi(\omega)$ of the $\chi(\omega)$ renormalization value; d. Renormalized with the electron-phonon interaction the imaginary part $\text{Im}\chi(\omega)$ of the $\chi(\omega)$ renormalization value. Frequency $\omega$ is expressed in dimensionless units (as a fraction of the maximum frequency component 0.275 eV of the phonon spectrum for this hydrogen sulfide phase). All results were obtained for the three pressure values P = 150 GPa ($\lambda$ = 1.001), P = 180 GPa ($\lambda$ = 1.11), P = 225 GPa ($\lambda$ = 0.92) and the temperature T = 200K.

On the Fig. 11, unlike the situation for the $SH_3$ phase, alternating behavior of $\text{Im}Z(\omega)$ at low frequencies $\omega < 2$, and hence the appearance of a set of "pockets", which are located at the distances of energy within the order of unity from the Fermi level is practically not observed.

*7*. RESUME

The calculations of the properties of the stable $SH_3$ orthorhombic sulfide structure IM-3M for the three pressure values P = 170 GPa, P = 180 GPa, P = 225 GPa and for the $SH_2$ structure with the symmetry I4 / MMM (D4H-17) for the three values of pressure P = 150 GPa , P = 180 GPa, P = 225 GPa at the temperature of T = 200K are performed. The renormalized frequency dependence of the real $\text{Re}\Sigma(\omega)$ and the imaginary $\text{Im}\Sigma(\omega)$ part of the self-energy part $\Sigma(\omega)$ of the electron Green's function, the real $\text{Re}Z(\omega)$ and the imaginary $\text{Im}Z(\omega)$ part of the complex renormalization of the electron mass, the real $\text{Re}\chi(\omega)$ and the imaginary $\text{Im}\chi(\omega)$ part of a complex $\chi(\omega)$ value, as well as the renormalized with the strong electron-phonon interaction electron density of states $N(\varepsilon)$ in the $SH_3$ and $SH_2$ phases of hydrogen sulfide are obtained. Renormalized density of electronic states differ from the "bare" electronic spectrum, which takes into account only the exchange and correlation effects of the electron-electron interaction in the ideal crystal lattice [3–9] excluding the strong electron-phonon interaction. The calculations used realistic calculated with the high accuracy "bare" electron and phonon characteristics of $SH_3$ and $SH_2$ hydrogen sulfide phases for the corresponding pressures [3,4]. Self consistent accounting of the variability of the electron density of states in the e-band, while taking account of strong electron-phonon interaction leads to the possibility of electron-phonon transitions in the substantial width of the conduction band, unlike usually the case, in quantum transitions of electrons within the layer with the thickness $\omega_D$ at the Fermi surface. Renormalized density of electronic states differ from the "bare" electron spectrum which takes into account only the exchange and correlation of the electron-electron interaction in the ideal crystal lattice [3–9] excluding the strong electron-phonon interaction.

$SH_3$ phase is characterized by a much more strong electron-phonon interaction and a lesser energy of the phonon frequencies compared to $SH_2$ phase. In addition, in the $SH_3$ phase the renormalized electron density of states adjacent to the Fermi level has sharp peaks (Van Hove singularities) compared to a smooth behavior of the electronic density of states in the $SH_2$ phase. In the $SH_3$ phase unlike the $SH_2$ phase, the effect is further detected of the reconstruction renormalized with the strong EP interaction conduction band (Fig. 6) with the appearance of a series of narrow not energy overlapped "pockets" which together constitute the reconstructed portion of the conduction band. These pockets are visible on Fig.7.b. on an example of ImZ behavior versus frequency. These factors indicate that the studied $SH_3$ phase is more promising superconducting material than the $SH_2$ phase. With increasing pressure the behaviors of the real $\mathrm{Re}\,Z(\omega)$ and the imaginary $\mathrm{Im}\,Z(\omega)$ parts of the complex renormalization of the electron mass, the real $\text{Re}\,\chi(\omega)$ and imaginary $\text{Im}\,\chi(\omega)$ parts of the complex $\chi(\omega)$ value are smoothed. The height of the renormalized peak of the electron density of states has been virtually unchanged, while the width of the peak decreases. Such changes to the characteristics of the hydrogen sulfide with pressure may be summarily described as an adverse effect of pressure on the superconductivity and may lead to a conclusion about the prospects of research in the field of high-temperature properties at the lower pressures. As a result of the self consistent consideration of the renormalization of the electron spectrum in the hydrogen sulphide SH$_3$ phase with the strong electron-phonon interaction, we found that hydrogen sulfide in a SH$_3$ phase at a pressure of P~225 ГПа and a temperature of T = 200K is a metal with a strong non-adiabatic effects. The SH$_3$ phase characteristic phonon frequency is of the same order of magnitude with the characteristic width of the energy pockets of the reconstructed electronic conduction band renormalized with the strong $\lambda \approx 2.2$ electron-phonon interaction, so that $\hbar\omega_D \sim E_{cond}$ in each of these pockets.

Thus, the SH$_3$ phase with IM-3M hydrogen sulfide lattice symmetry is one of the two main candidates responsible for the superconducting properties of the metallic hydrogen sulfide. The SH$_3$ phase with the IM-3M hydrogen sulfide lattice symmetry is a substance, the calculation of which the superconducting properties should be carried out by generalized Éliashberg theory methods that take into account the finiteness of the electronic band gap. Superconductivity in this phase of the hydrogen sulphide can not be treated with the standard solutions for the Éliashberg equations [16 – 24] corresponding to the infinite width of the electron band which do not take into account the abrupt variations of the "bare" electron density of states. When considering superconductivity a strong renormalization of the electronic density of states with phonons should be considered , leading to a sharp renormalization of both width and qualitative structure of the conduction band. The normal state of the hydrogen sulfide and the superconductivity in this matter should be considered by a generalized Éliashberg theory [10,11], originally developed for the consideration of normal and superconducting properties of the cuprates, but totally applicable to the consideration of the properties of other substances with electron band of finite width, strong electron-phonon interaction, electron- hole nonequivalence.

The authors thank N.N. Degtyarenko for providing the results of the calculations of the "bare" electron and phonon spectra of the hydrogen sulfide phases.
The study was supported by a grant from the Russian Science Foundation (project №14-11-00258).